\documentclass[runningheads,a4paper]{llncs}

\usepackage{amssymb}
\usepackage{graphicx}

\usepackage{subfigure}

\usepackage{url}
\urldef{\mailsa}\path|{emmenual.reddy, nicholas.rollings, rohitash.chandra}@usp.ac.fj|
    
\newcommand{\keywords}[1]{\par\addvspace\baselineskip
\noindent\keywordname\enspace\ignorespaces#1}

\begin{document}

\mainmatter  

\title{Mobile Application for Dengue Fever Monitoring and Tracking via GPS: Case Study for Fiji}

\titlerunning{Mobile Application }

%
%
\author{Emmenual Reddy \inst{1,3} \and Sarnil Kumar \inst{1,3} \and Nicholas Rollings \inst{2,3}  \and Rohitash Chandra \inst{1,3}}
\authorrunning{Mobile Application }

\institute{ School of Computing, Information and Mathematical Statistics,\\The University of the South Pacific, 
Laucala Campus, Suva, Fiji. \and Geography, Earth Science and Environment, \\The University of the South Pacific, 
Laucala Campus, Suva, Fiji.  \and Artificial Intelligence and Cybernetics Research Group, Software Foundation, Nausori, Fiji\\
\mailsa\\
\url{http://www.usp.ac.fj}}

%
%

\toctitle{Lecture Notes in Computer Science}
\tocauthor{Authors' Instructions}
\maketitle

\begin{abstract}
The 2013 outbreak of Dengue in Fiji resulted in an alarming number of deaths and has been is a matter of serious concern. Dengue fever is a disease caused by the four types of the Dengue virus serotypes  and transmitted mostly from mosquito bites. In Fiji, dengue diagnosis is only done in hospitals which are slow and time consuming. It is also important to monitor the spread of Dengue. Fiji needs an convenient method of monitoring the spread of Dengue. With increase in affordable smartphones and better Internet coverage, there is scope for  a mobile application for Dengue fever monitoring and tracking. This paper proposes  a mobile application for Dengue monitoring based on global positioning system (GPS) enabled mobile phone technology. It also provides an information network that shows the spread of dengue which will allow health authorities to quickly identify dengue infected areas in Fiji. A mobile application prototype is developed and tested and the scope for further testing and implementation is also given.

\keywords{Dengue, Dengue virus, Dengue monitoring, Dengue tracking, GPS, Mobile application}
\end{abstract}

\section{Introduction}

Health information technology, which includes computers, mobile devices and other devices used in the management of medical information, has great potential to promote health and support health care around the world  \cite{Chan2010300}, \cite{Williams2008503}, \cite{Agha201419} \cite{Haux2006795}, \cite{Zuckerman2009965}. Mobile phones are more preferred in the delivery of health information technology because of its widespread adaption and its increasing processing power.People tend to carry mobile phones with them everywhere and unlike desktop computer, mobile phone are nearly always with the person \cite{Klasnja2012184} \cite{Thomas201269}.

In developing countries there are many challenges faced when designing and implementing ways to improve the health information technology. The problems faced are limited health care resources, access to technology infrastructure and technical expertise to support technology \cite{Lluch2011849} \cite{Belanger2012654}. Other difficulties faced by people in developing countries when using mobile phone are Internet connectivity, mobile network signal and access to electricity and charges \cite{Zafar2014236}.

There are challenges also faced when deploying the health information technology in the healthcare workforce. One major challenge is the lack of knowledge in information technology. Illiteracy itself may be one of the central barriers that could prevent the effective use of these phones in this situation; Illiteracy is often associated with poverty, which coupled with the inability to read may limit an individual's access and ability to benefit from such technologies \cite{Chudgar201420}. Healthcare workforce is often scarce and overburdened already with responsibilities. Technology interventions in the health care should not require high learning curves, it should be flexible and easily understood so it will not consume their time \cite{Chan2010300}.

According to \cite{Chawla2014169}, traditional medicine offers an alternative solution and could be explored as a safer treatment option. One such treatment includes using the Papaya leaf remedy to cure dengue which has been proven to be successful in a research carried out in Pakistan \cite{Ahmad2011330}.

Dengue fever is a major health concern in Fiji with a number of dengue outbreaks occurring in the last three decades  \cite{Tabua2013}. The most recent outbreak was declared in December 2013 \cite{Vuibau2013}. In the latest outbreak there were over 446 suspected cases of Dengue notified to the Ministry of Health and there have been 12 confirmed deaths \cite{Nand2014}. Dengue fever is a disease caused by any of the four types of the dengue virus serotypes (DEN 1, DEN 2, DEN 3, DEN 4) \cite{WHO2009} and transmitted mostly from mosquito bites \cite{Muller201290}.  

Traditionally, the diagnosis of dengue infection is a clinical determination by a medical professional. However, during the early infection period, when symptoms may not be as severe as later in the course, the patient with dengue fever can appear to have a less serious illness and therefore may not be given appropriate treatment \cite {Matthews:2012}. The early recognition of dengue symptom is very important, however, dengue detection kits are only accessible in hospitals because of the high costs \cite{Mitra2014450}. Therefore, a more convenient method of diagnosing dengue in patients is needed.

The Global Positioning System (GPS) is a satellite-based navigational system that was developed by the United States  Department of Defence in the early 1980s \cite{Wang2013202}. The GPS technology allows the accurate positioning of an object using satellite signals \cite{Mintsis2004399}. GPS technology has been  used in many health applications \cite{Kerr2011} \cite{Rainham2008} \cite{Boulos2011} \cite{Chaix2013} .

In this paper, a mobile based dengue fever monitoring and tracking application is proposed that monitors the spread of dengue via global positioning system (GPS). The application allows likely dengue patients to enter their symptoms and receive immediate vital feedback. The application provides users with advice on how to prevent dengue fever. The application is informational and not a substitute for visiting a doctor.  In the case of detected or diagnosed  fever or any related symptoms, the users are advised to immediately see a physician. The users’ current location which is tracked through the GPS tracker and other important user information is stored in a database that can be further be integrated with existing electronic medical record systems. Using the data collected, a map showing the locations of likely dengue infected people is then produced.  A mobile application prototype is developed and tested and the scope for further testing and implementation is also given. Relevant authorities such as Ministry of Health or the town councils can   use this information to identify dengue outbreak areas in Fiji and precautionary measures can be taken.

The rest of the paper is organised as follows. Section 2 gives a background of the area. Section 3 presents the proposed design and Section 4 shows Implementation and results. Section 5 concludes the paper with discussion of future work. 

\section{Background}
According to the World Health Organisation, primary prevention is the most effective measure in dengue prevention and control since no vaccine is currently available \cite{WHO2009}.

There is a dire need for a dengue vaccine to further prevent the spread   and contain the growing pandemic. Several vaccine companies and research groups are actively pursuing the development of a vaccine to prevent dengue \cite{Cassetti20143115}. While a licensed dengue vaccine is not yet available, several vaccine candidates are currently being evaluated in clinical trials \cite{Schmitz20117276}.

Since no effective vaccine is available for treating dengue, the eradication or control of the main mosquito vector is regarded as essential. Research has also been done in devising control methods for the \textit{Aedes Aegypti} mosquitoes which is the main type that spreads dengue virus \cite{Morrison2008}. 

In developing   countries, viral outbreak of Dengue causes a heavy financial constrain in terms of manpower (medical staff) and resources such as hospitals and equipment \cite{Humayoun2010e54}. In addition,  there is little awareness  about the symptoms of Dengue fever.

A number of studies have been conducted to determine the usefulness of herbal medicine in curing dengue. Researches have indicated that the juice of the leaves of the Carica papaya plant from the family Caricaceae could help to increase the platelet levels in these patients. \cite{Subenthiran2013} \cite{Senthilvel2013} \cite{Sarala2014}

The World Health Organization has estimated that out of the 2.5 billion world population who are currently at risk of suffering dengue in over 100 countries, two-thirds reside in the Asia Pacific Region \cite{WHO2009}.
Fiji, a multi-cultural south pacific island nation with a population of around 800,000, had its share of dengue fever outbreaks in the past. Major dengue outbreaks occurred in 1998 and 2003 that resulted in loss of lives, substantial health care stress and considerable economic costs \cite{Tabua2013}. The latest dengue outbreak was declared in December 2013 \cite{Vuibau2013}.

Electronic Medical Records (EMR) are used in medical sector in various ways which includes easy access to medical record, reduction of potential medical error, appointments  and billing management \cite{Park2012204} However, there are obstacles to a successful adoption of electronic records and to the transition from a paper-based paradigm. These obstacles are not only technological, but often also related to the deployment procedures followed \cite{Oliveira2013} \cite{Scholl2011}.  The purpose of EMR  system is to have a centralized location of all information regarding a patient which can be shared by doctors, hospitals and patients. \cite{Hatton2012706}.  

Fiji has used PATIS (Patient Information System) which was developed with AusAid funding for the Pacific nations for managing patient records. Initially, PATIS was still paper based and not web based. PATIS is a low-cost application software system that provides functionality generally found in commercial patient administration systems modules of a Health Information System. There have been significant enhancements to Fiji's PATIS, which now has functionality usually only provided by specialised clinical system modules \cite{Soar2012}. A review team in 2010 was informed that the PATIS system is being transferred to one that is web based and this will help encourage wider use of the very valuable data being generated by the system \cite{Freeman2010}. The latest version of PATIS with improved accessibility and speed is referred to as PATISPlus \cite{FijiMOH2014}.

OpenEMR is an open source software used in health information systems \cite{Feufel2011e85} \cite{Ford2006106}. Open source software solutions are becoming popular in developing countries as they can be deployed taking into consideration the cost of purchasing licensed software \cite{Kiah2014} \cite{WikipediaOpenSource}.

\section{Proposed System}

The motivation to develop a mobile based dengue monitoring system came from the alarming rate of dengue deaths in the recent dengue outbreak in Fiji \cite{Sauvakacolo2014}. There is a need for a faster and convenient method of checking for possibility of dengue infection.   

We propose a mobile application that can be used by mobile devices such as  smartphones and Tablets as these devices are becoming  more affordable smartphones and there is better network coverage in Fiji when compared to past years.

During an outbreak of Dengue in Fiji, the proposed mobile application can be used by the Ministry of Health  to monitor the spread of the disease and take precautionary measures by identifying affected areas and provide support with help of the media. The proposed mobile application enables a symptoms check of Dengue and in order to take the precautionary steps that is advised by the mobile application and can reduce a significant amount of workload of the medical practitioners. 

There is no vaccine cure for dengue, treatment consist of early identification of the disease with intensive surveillance and fluid support \cite{Matthews:2012}. 
The symptoms of the dengue which will be assessed in this mobile application are fever, severe headache, severe eye pain, joint and muscle pain(myalgia), skin rash, mild bleeding manifestation, low white blood cell count, nausea/vomiting and joint swell \cite{Humayoun2010e54} \cite{WHO2009}.

There are some limitations in using mobile application since not everyone has a smartphone. Some people are not smartphone literate, meaning they do not know how to install and use  mobile applications. In addition, using the application requires the user to have internet credit in his smart phone and also to have the mobile data enabled. Despite this limitations, the proposed system has many advantages as it can be used to identify and control the spread of dengue. The proposed system's features are discussed in the following subsections. 

\begin{figure}
    \begin{center}
    \includegraphics[scale = 0.45]{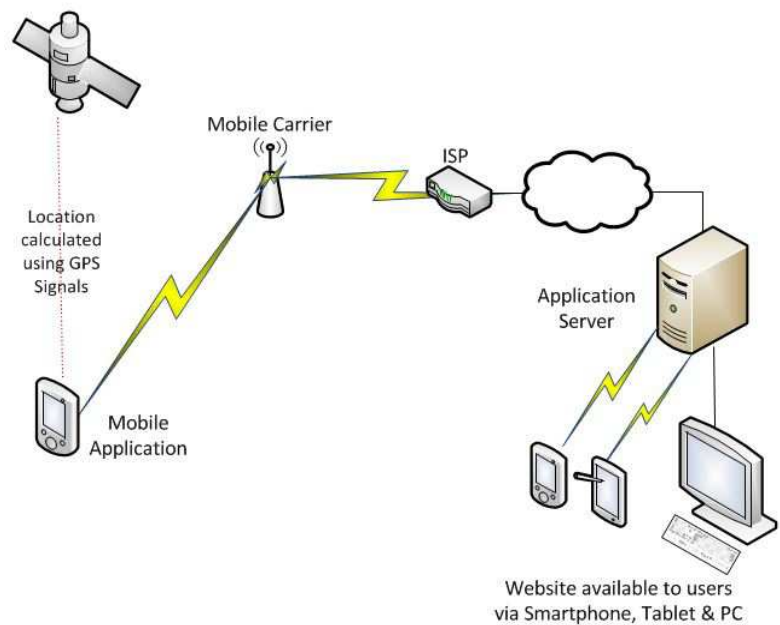}
    \caption{Overall architecture of the system showing the path of data transmission from the mobile application through the mobile carrier and the Internet Service Provider (ISP) to the application server and finally to one of the computing devices.}
  \end{center}
\end{figure}

\subsection{Overview}
The objectives of the dengue fever monitoring and tracking application are to provide the general public with a dengue symptoms checker, receive feedback of the papaya leaf remedy \cite{Ahmad2011330}, self-report a Dengue case and to store the location of the user via GPS. The data (GPS coordinates) which is stored in the server will be featured in the website showing the location of users infected by Dengue.

\begin{figure}[htb!]
    \begin{center}
    \includegraphics[scale = 0.45]{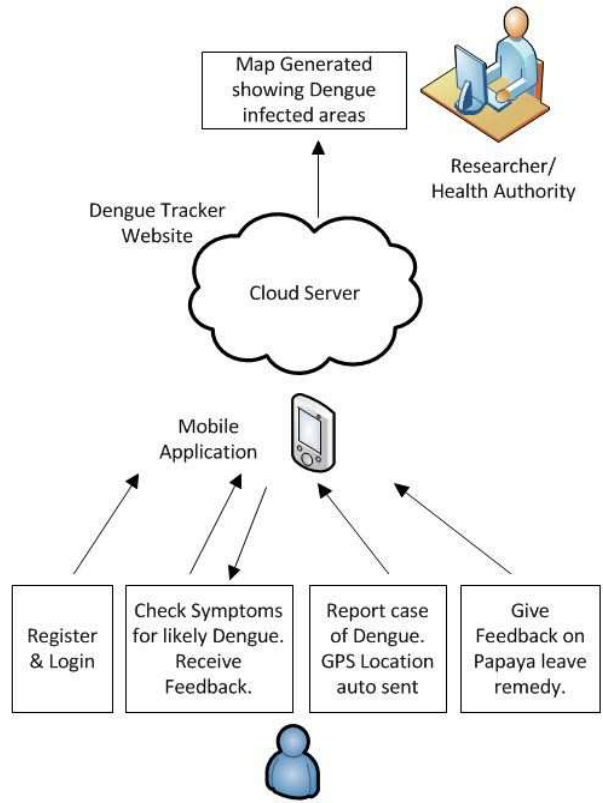}
    \caption{The proposed mobile application system showing the different features of the system including the features to register and login, checking symptoms for dengue, reporting a confirmed case of dengue and giving papaya leaves feedback. }
  \end{center}
\end{figure}

\subsection{Authentication}
The access to the dengue symptoms monitoring and tracking application is given to registered users. The mobile application has a feature where the general public can register and login to the system. In the registration process, the user is asked the following details; name, email address as username, password, date of birth, gender, height and weight. To login to the application, the user needs to enter their email address and password.

\subsection{Dengue Symptoms Monitoring}
Once the user has successfully logged in the application, they can use the \textit{ Dengue Systems Check} feature to verify their symptoms for possibility of dengue infection. The mobile application determines the likelihood of the users being infected with dengue based on their symptoms. If the users have selected fever and any two other symptoms from the eight other symptoms, they are informed that they are highly likely to be infected with Dengue fever \cite{WHO2009}. They are also advised to visit a physician immediately.

\subsection{Dengue Case Reporting and Tracking}
As discussed earlier, there is no vaccine cure for dengue but there are ways in which we can control dengue outbreak. One way to control dengue is through early detection and surveillance \cite{Muller201290} \cite{Cleton2014159}. The dengue surveillance allows public health workers in resources-limited areas to identity areas with high Dengue infections and potential larval potential larval development areas such as garbage piles and large pools of standing water. \cite{Chang2009}. 

In this mobile application, there is a feature where the patient can report a case of Dengue. Once the user reports a case, the location of the user is updated on the database. The location of the user is  based   in terms of GPS coordinates; latitude and longitude referenced to WGS84.

The locations of the likely dengue infected people are used to generate a Dengue map that is given on the Dengue tracker website. Relevant authorities such as the Ministry of Health in Fiji and town councils can use the information on the website to track the possible dengue outbreak and mosquito breeding areas and take appropriate action. This deployment in mobile application can be a very efficient and cost effective way of controlling the spread of Dengue \cite{Chang2009}.

\subsection{Feedback on Papaya Leaves}

Dengue fever alternative medical treatment can be done with papaya leaf extracts \cite{Sarala2014}. Studies have shown that platelet count increases by taking papaya leaves extracts \cite{Subenthiran2013} \cite{Senthilvel2013}. In a case study in Pakistan, a 45 year old man infected with dengue virus who was given Carica papaya leaves extracts every morning and afternoon noted an increase in platelet counts in the blood cells \cite{Ahmad2011330}.

In Fiji, dengue infected patients have also tried papaya leaf remedy, however, it is not known how effective it has been in terms of platelet counts. The proposed mobile application ensures that such alternative medicine treatment results are deposited in the system for further research. The  feature allow users who have tried papaya leaf to provide feedback by checking their platelet counts through blood test by doctor  before and after the papaya leaf treatment and feeding it into the mobile application. For example, if a person visits a doctor and after examination it is found out that the patient has dengue fever with low platelet counts. The same patient takes the advice of taking papaya leaf remedy. On the patient’s next medical examination if the platelets count increases, the users can give feedback via mobile application.

This would assist relevant authorities such as researchers and the Ministry of Health officials  to determine whether papaya leaf remedy can really help  Dengue infected patients in Fiji.  The feedback can be also used for further research in drug discovery.  If it works, it would be a breakthrough in Dengue research as currently there is no effective vaccine to cure Dengue fever. It would also greatly assist in tackling Dengue cases and saving lives with a very cost effective way.

\subsection{Website}
The website features data which is uploaded on the database from the users of the mobile application. On the website,  a map can be generated using the GPS coordinates from the database. This will show markers of the location of users who have reported through the mobile application who have been infected with Dengue. When a researcher zooms in to a particular marker on the map, they will be able to see the exact location of the infected users. Moreover, a particular area concentrated with many markers would suggest that it could be possible breeding grounds for mosquitoes. Health authorities can use this vital information to quickly identify areas in Fiji which needs to be fumigated and for health workers to visit people living in this areas to control further spread of Dengue.

\begin{figure}[htb!]
    \begin{center}
    \includegraphics[scale = 0.27]{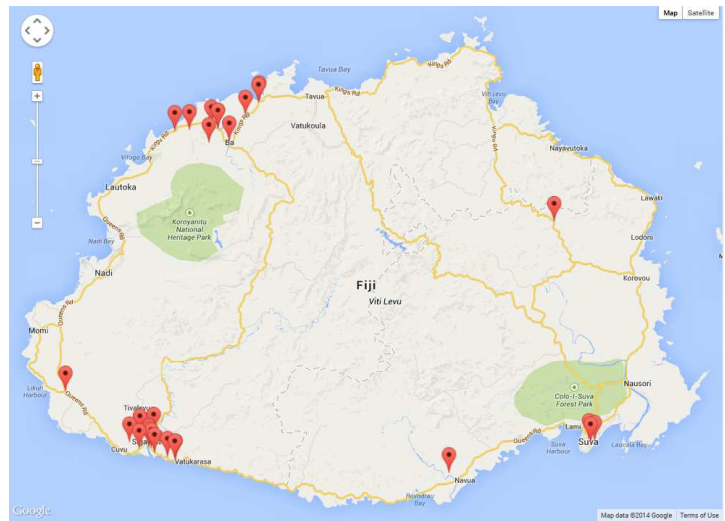}
    \caption{Map showing markers which indicates the location of people infected with Dengue Fever in the island of Viti Levu, Fiji.}
  \end{center}
\end{figure}

\begin{figure}[htb!]
    \begin{center}
    \includegraphics[scale = 0.27]{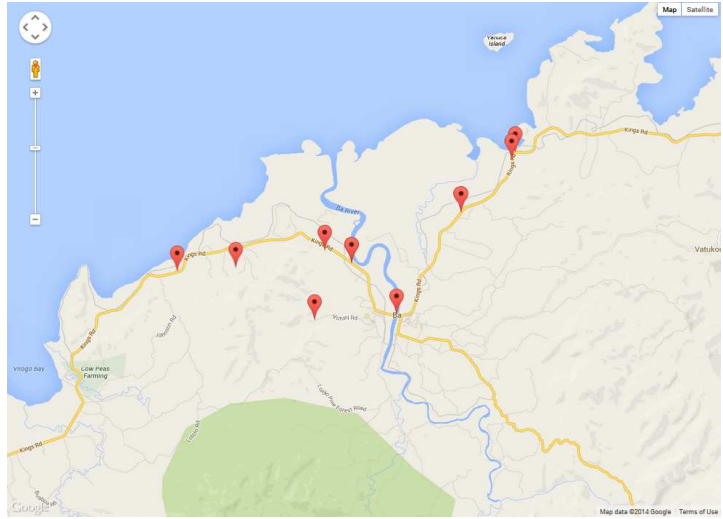}
    \caption{Map when zoomed in showing markers which indicates the location of people infected with Dengue Fever in the area of Ba, Fiji.}
  \end{center}
\end{figure}


\section{Implementation and Results}
In this section, we provide details of implementing the proposed application built on an Android platform. Firstly, the implementation details are discussed which includes the selected programming language and other supporting software required to develop the mobile application. Global Positioning System (GPS) is also discussed as it is a key feature used in the mobile application to determine the location of the user. Finally, details of testing the mobile application are discussed.

\subsection{Implementation Details}
The Android application is developed using Java programming language. Java is a class-based object orientated language and is one of the most popular programming languages in client-server web application \cite{TechHuddle}. Once Java program is complied, it can run on any platform regardless of the computer architecture \cite{OracleJava}. The Android SDK (Software Development Kit) provides the API (Application Program Interface) libraries and developer tools necessary to build, test, and debug applications for Android \cite{Android}.

The data sent from the application is stored on a LAMP (Linux Apache MySQL PHP) server \cite{Linus2014}. The data is then sent in JSON\footnote{JavaScript Object Notation} format \cite{Json2014} from the android application and updated on the database using PHP\footnote{Hypertext Preprocessor - PHP is a server-side scripting language designed for web development but also used as general purpose programming language} and MySQL\footnote{My Structured Query Language - MySQL is an open source relational database management system (RDBMS) based on Structured Query Language}  \cite{Oracle2014}.

The website\footnote{http://www.denguetracker.com} is designed in WordPress \cite{Wordpress2014} which is a content management system built on PHP and MySQL. Since WordPress also uses PHP and MySQL, it is easy to get data from the database and display it on the website.

\subsection{Global Positioning Systems (GPS)}
GPS allows devices to track the position of individuals in latitude and longitude along with timestamps. The limitation of the GPS is that signals are usually blocked in indoor or underground places, GPS devices get interference from tall buildings and continuously collecting GPS data normally consume devices energy quickly \cite{Lin2013}.

The in-built GPS receiver in the smartphone is used to get the GPS coordinates of the users who reports to be likely infected with dengue fever. The GPS coordinates are stored in the server and then used to generate the map showing the areas where people infected with dengue are located.

\begin{figure*}[htb!]
  \begin{center}
    \begin{tabular}{c}  
       \subfigure[ Functions of the mobile application.  ]{ \includegraphics[width=1.7in, height=3in ]{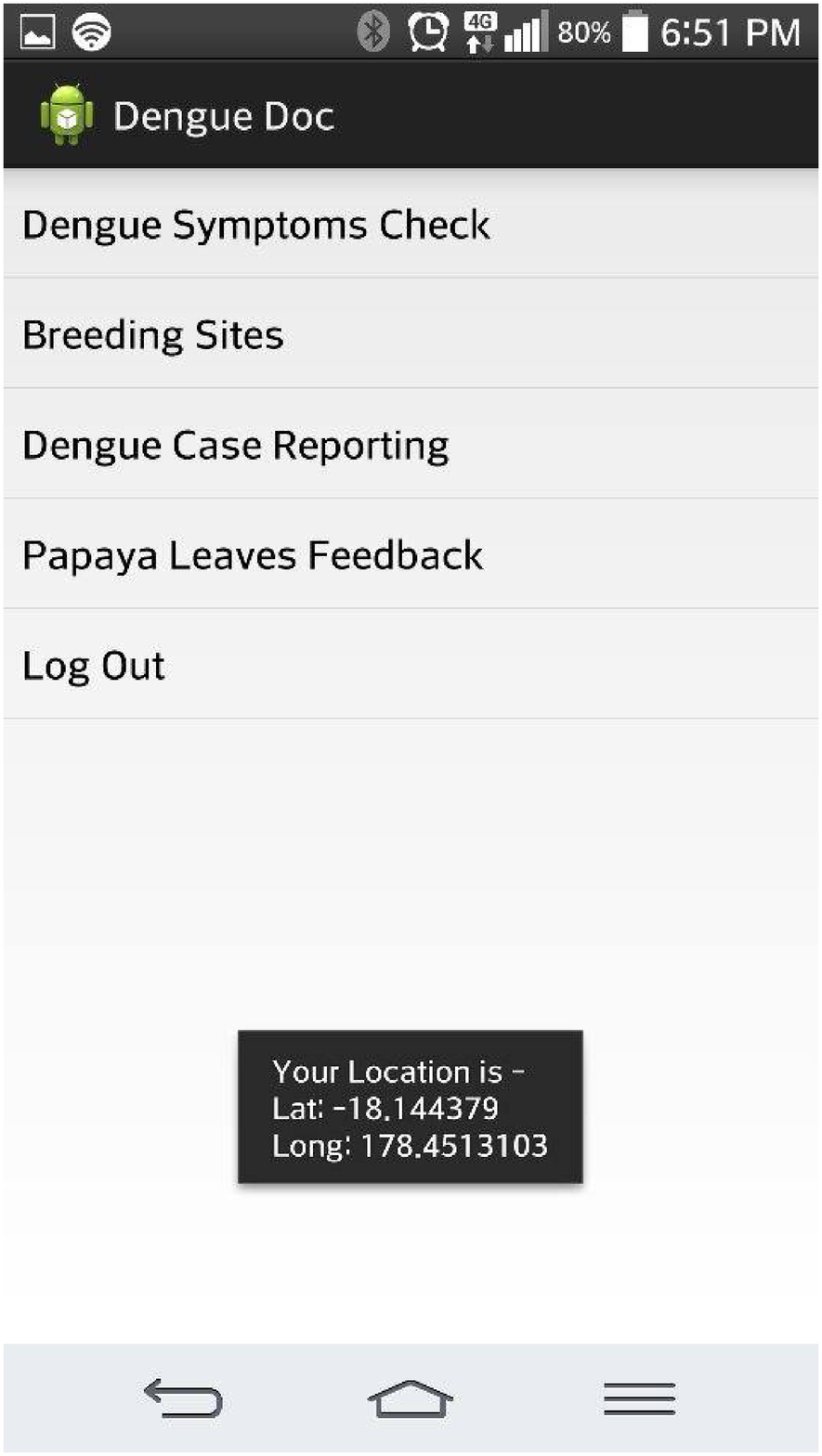}} 
    \subfigure[ Feature that allow users to select symptoms and check whether they are likely infected with Dengue Fever.]{ \includegraphics[width=1.7in, height=3in ]{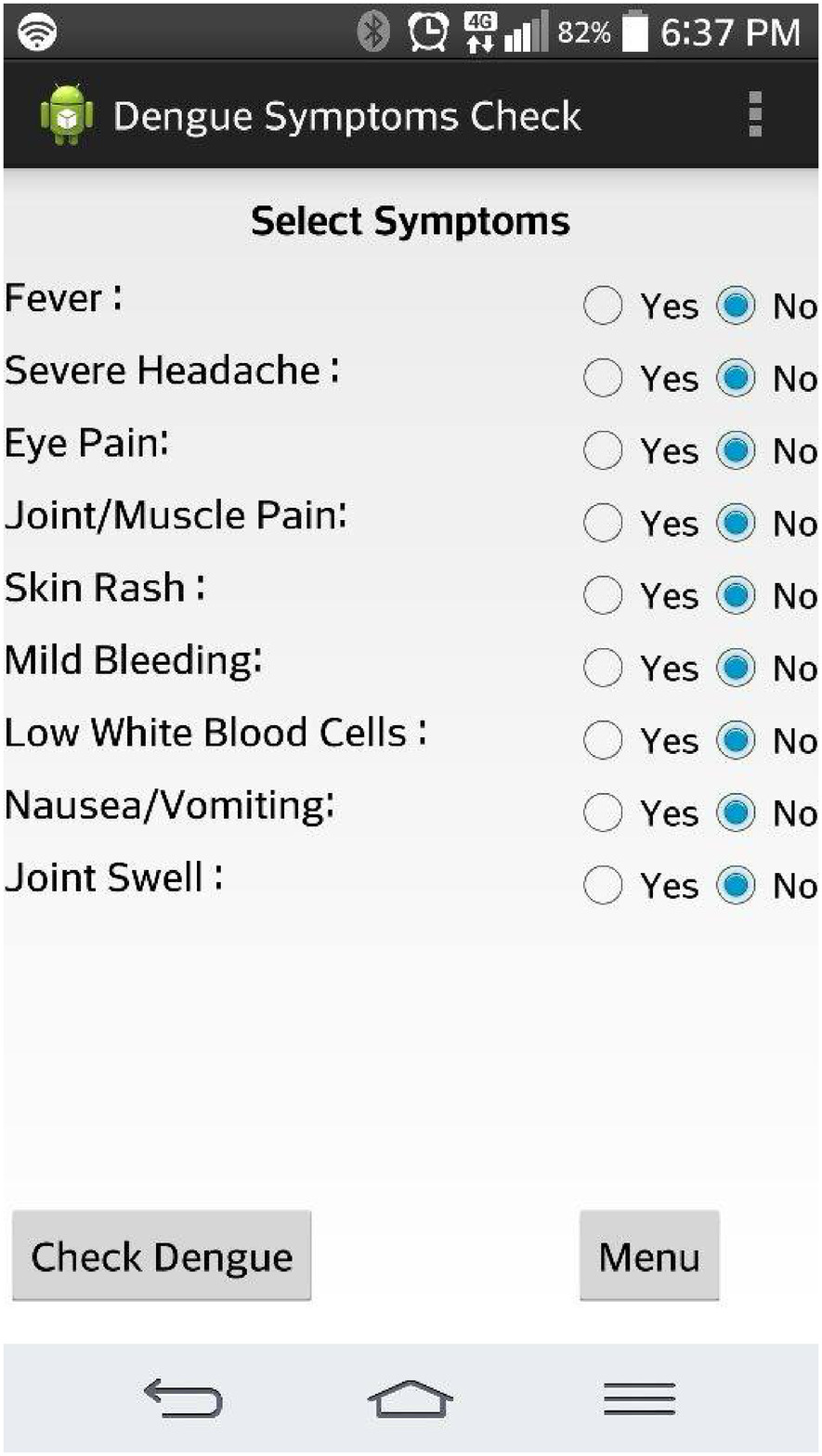}}\\ 
    \end{tabular}
    \caption{The mobile application provides a menu of functions once the user has successfully logged in. A user may select the first option to carry out dengue symptoms check.}  
 \label{fig:sunspotplot}
  \end{center}
\end{figure*}

\begin{figure*}[htb!]
  \begin{center}
    \begin{tabular}{c}  
       \subfigure[ Feature that allow users to report a confirmed case of dengue.  ]{ \includegraphics[width=1.7in, height=3in ]{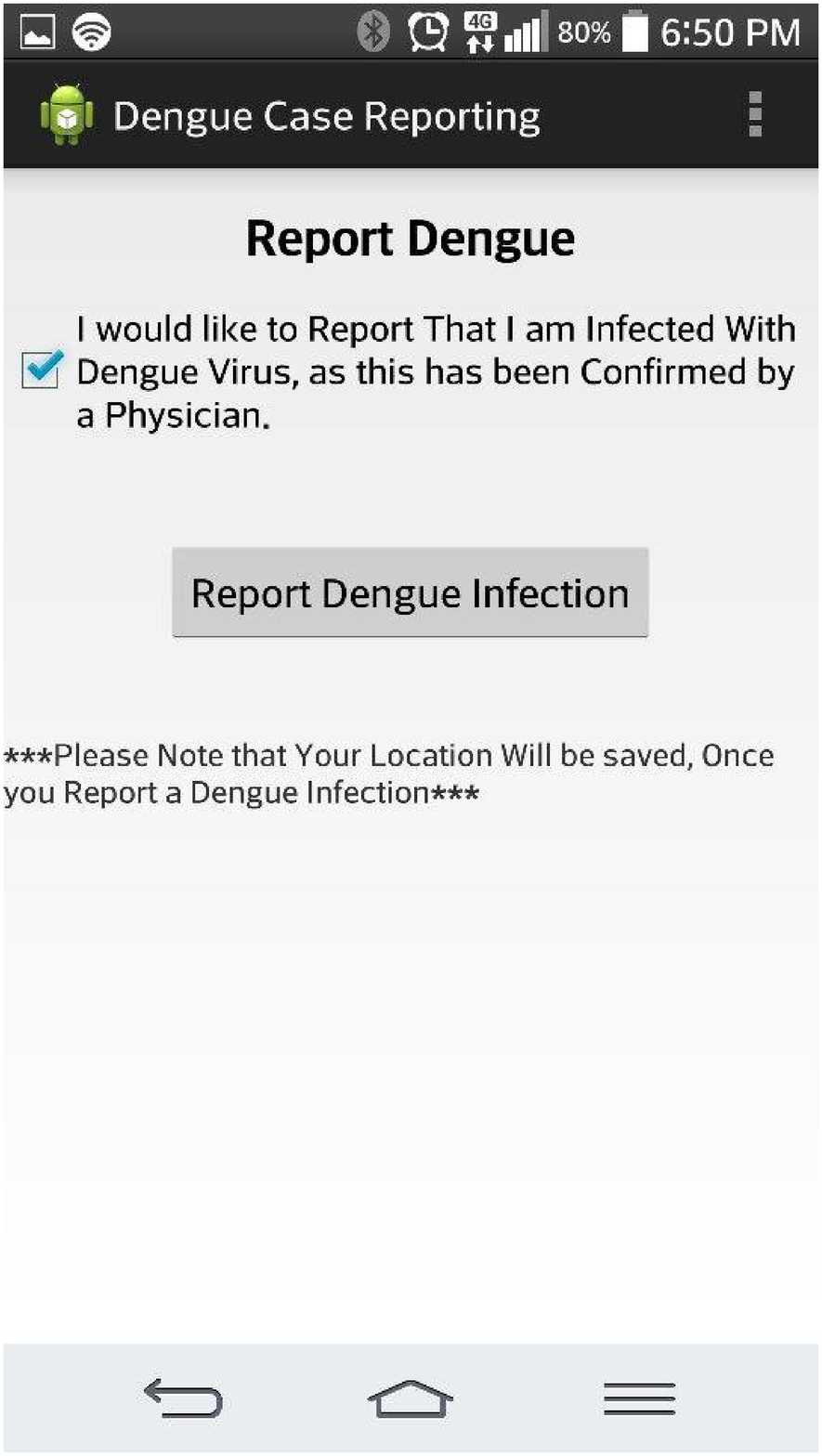}} 
    \subfigure[Feature that allow users to give feedback on the use of Papaya Leaves Remedy.]{ \includegraphics[width=1.7in, height=3in ]{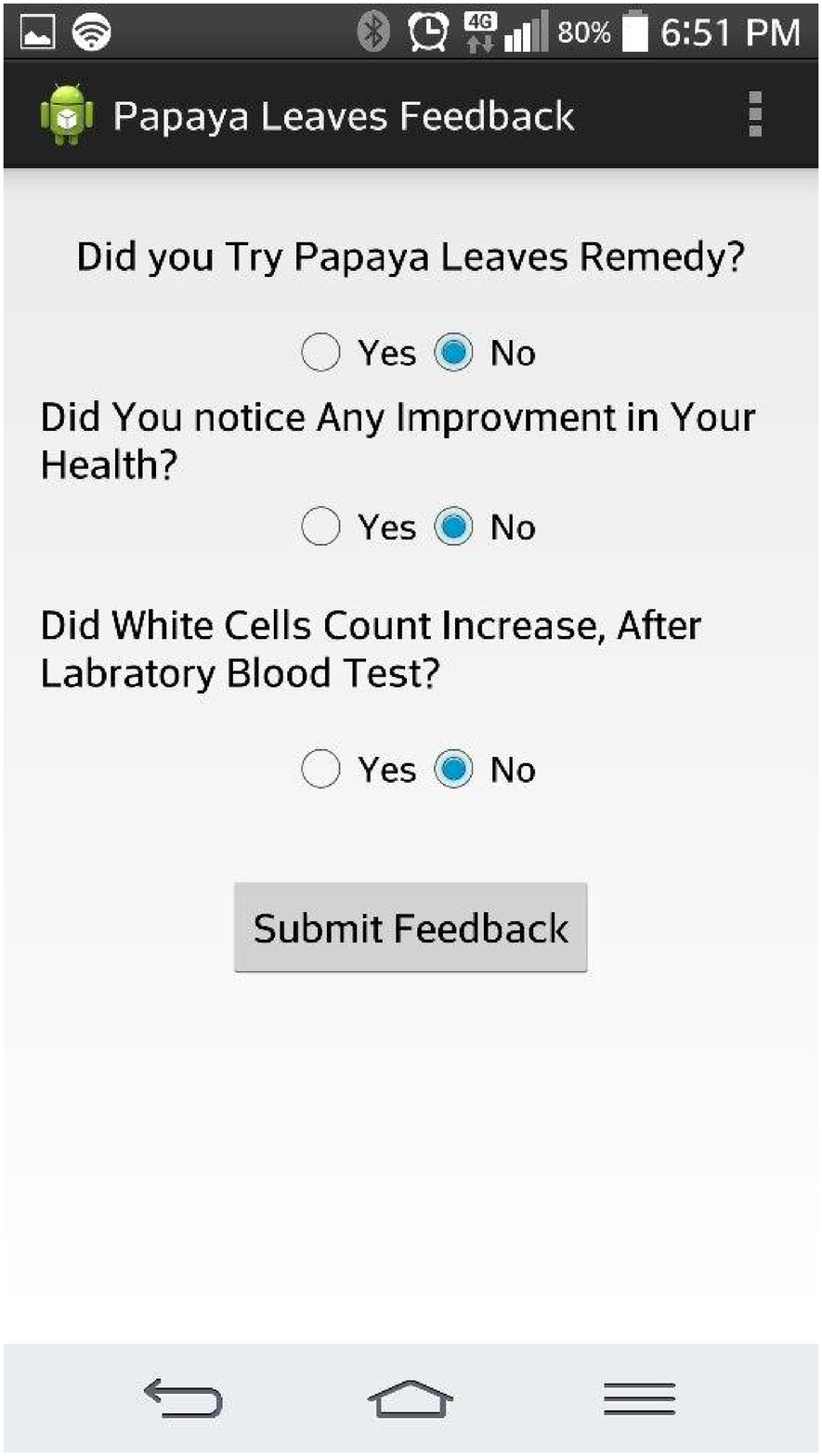}}\\ 
    \end{tabular}
    \caption{The mobile application also provides features to report a confirmed case of dengue and to give feedback on the use of the Papaya Leaves Remedy.}  
 \label{fig:sunspotplot}
  \end{center}
\end{figure*}  

\subsection{Testing the Mobile Application}

Testing is vital  as it determines the features of the application that can be reliable and useful  under heavy load and features that can be accessible with limited or no Internet \cite{KeynoteMobile}. The mobile application was made available to 10 android smartphone users for testing. Users were asked to perform the role of someone who is suffering from Dengue. Following the trial, the users were interviewed regarding their experience. Users in general,  gave positive feedback for user performance and satisfaction in areas with good Internet connectivity. However, they indicated that they were not able to register and log in to the mobile application when the internet connectivity was poor.

The application needs further improvement  and wider testing to evaluate system performance which can be done in future work. There are many challenges in mobile application testing including variety of mobile devices, diversity in mobile platforms, mobile network operators and scripting. Server load testing can be performed in future work once the application is implemented for use by more than 100 users at one particular time.

\section{Discussion}
The mobile based dengue fever monitoring and tracking application is developed that allow likely dengue patients to enter their symptoms and receive immediate vital feedback. The application provides users with advice on how to prevent dengue fever and basic treatment. The application for is health information only and not a substitute for visiting a doctor. In the case of fever or any two other related symptoms, the users are advised to immediately see a physician. The users current location which is tracked through the GPS tracker and other important user information is stored in a database. Using the data collected, a map is generated showing dengue infected areas. Relevant authorities such as the Ministry of Health and town councils can use this information to identify dengue outbreak areas in Fiji and notify their workers to visit these areas.  One potential problem with this tracking system is the patient may realise the symptoms of dengue when they are some distance from the actual infection point. This will give a false impression of where the patient was actually infected. In future editions it is suggested that during peak Dengue season users movements are recorded hourly for the previous 20 days.  After 20 days the older positions are overwritten allowing fro a balance between the amount of data collected and ensuring the incubation period for Dengue is covered. If multiple infected persons have visited the same geographic location it raises the possibility this is a potential site of infection.

There are several important implications on how this would affect the monitoring and control of Dengue. Due to the portable nature of the smartphones and Tablets, Dengue symptoms check can be done at the patient’s home. The mobile based symptoms check aids in the identification of patients infected with dengue in rural areas. Subsequently, patients will be provided with advice on how to prevent dengue infection. The mobile symptoms checkup will have the largest impact on children who are the most susceptible to the serious side effects of dengue.

One of the drawbacks of the dengue application prototype is the application platform. The dengue application is created using Android platform, therefore, it will only work in an Android based devices. Hence, it will not work in an iOS\footnote{iPhone OS is a mobile operating system developed by Apple Inc.} or windows platform based smartphones and Tablets. However, the Android application can be developed in other platforms using the proposed system design so that it could work in all types of phones. Secondly, the application will only work if the user has mobile data enabled. Without enough mobile credit, a user will not be able to activate mobile data and therefore will not be able to use the application.

The system that is developed is not going to replace physicians, however, it will greatly assist them in making their work easier in controlling disease outbreaks.

For mobile phones that do not have internet access, an alternative SMS based system can be used whereby users send their options and receive feedback through text messages. This alternative system can also be used on smartphones in case they do not have internet or when internet connectivity is poor. Other solutions include having wireless enabled in hospitals and health centres where patients can easily download and install the mobile application free of charge. In this case, their GPS location can be updated later and an option to do so can be given.

In future, if Ministry of Health in Fiji adopts OpenEMR for storing all patient data in electronic form, this Android application then can easily play an important role in updating issues related to dengue. The migration of the database will also be easy. 

\section{Conclusion and future work}
In this paper, we present a design of a Dengue monitoring and tracking system based on GPS. The proposed application provides an alternative and more convenient method for checking symptoms in likely dengue patients. GPS technology is also used to track the location of the dengue infected patients. Using the data that is collected, a dengue map is generated which helps health authorities to identify dengue infected areas. 

 The application can be improved to ensure real-time tracking of dengue outbreaks and compatibility of the application with iOS and windows mobile platform. To further enhance the capabilities of the application, features can be added that will be enable users to upload multimedia such as videos and photos  of Dengue breeding sites. The data can be then retrieved from the database and displayed on the Dengue tracker website and be used by the relevant authorities to clear and fumigate affected areas. 

Future work can involve  implementation of the application for large scale use of thousands of users and further testing to ensure that the features work when hundreds of users access the application simultaneously. Feedback from the relevant authorities through surveys can be used to  validate the system and improve it further. Data gathered can be published for further analysis by medical researchers and authorities that can build preventative systems.

\end{document}